\documentclass[11pt,fleqn,twoside]{article}
\usepackage{amsfonts,amssymb,latexsym}
\makeatletter
\newcommand{\prava}[1]{\small\it
\begin{flushleft}
Copyright \copyright \ 1999 by  #1
\end{flushleft}}

\newcommand{\name}[1]{\begin{flushleft}
                       \LARGE \bf #1
                       \end{flushleft}\vspace{-3mm}}

\newcommand{\Author}[1]{\begin{flushleft}
                       \it #1 \end{flushleft}}

\newcommand{\Adress}[1]{\begin{flushleft}
                       \it #1 \end{flushleft}}

\newcommand{\Date}[1]{\begin{flushleft}
                      \small  \it #1 \end{flushleft}}

\newcommand{\ehkol}{Author \ name}
\newcommand{\ohkol}{Article \ name}
\renewcommand{\@evenhead}{
\hspace*{-3pt}\raisebox{-15pt}[\headheight][0pt]{\vbox{\hbox to \textwidth 
{\thepage \hfil \ehkol}\vskip4pt \hrule}}}
\renewcommand{\@oddhead}{
\hspace*{-3pt}\raisebox{-15pt}[\headheight][0pt]{\vbox{\hbox to \textwidth 
{\ohkol \hfil \thepage}\vskip4pt\hrule}}}
\renewcommand{\@evenfoot}{}
\renewcommand{\@oddfoot}{}

\newcommand{\be}{\begin{equation}}
\newcommand{\ee}{\end{equation}}
\newcommand{\ba}{\hspace*{-5pt}\begin{array}}
\newcommand{\ea}{\end{array}}

\newcommand{\ds}{\displaystyle}
\makeatother

\begin{document}
\setcounter{page}{51}

\thispagestyle{empty}

\renewcommand{\ehkol}{P. Rudra}
\renewcommand{\ohkol}{Contact Symmetry of Time-Dependent
Schr\"{o}dinger Equation}

\begin{flushleft}
\footnotesize \sf
Journal of Nonlinear Mathematical Physics \qquad 1999, V.6, N~1,
\pageref{rudra-fp}--\pageref{rudra-lp}.
\hfill {\sc Article}
\end{flushleft}

\vspace{-5mm}

\renewcommand{\footnoterule}{}
{\renewcommand{\thefootnote}{} \footnote{\prava{P. Rudra}}

\name{Contact Symmetry of Time-Dependent Schr\"{o}dinger 
       Equation for a   Two-Particle System:  Symmetry 
Classif\/ication of Two-Body Central Potentials} \label{rudra-fp} 

\Author{P.~RUDRA}

\Adress{Department of Physics, University of Kalyani, Kalyani, WB,
741-235, India\\[1mm]
E-mail:  rudra@cmp.saha.ernet.in \ and \ rudra@klyuniv.ernet.in}

\Date{Received July 31, 1998; Accepted September 8, 1998}

\begin{abstract}
\noindent 
Symmetry classif\/ication of two-body central 
potentials in a two-particle Schr\"{o}dinger equation in terms 
of contact transformations of the equation has been 
investigated. Explicit calculation has shown that they are of 
the same four dif\/ferent classes as for the point
transformations. Thus in this problem contact transformations
are not essentially dif\/ferent from point transformations. 
We have also obtained the detailed algebraic structures of the
corresponding Lie algebras and the functional bases of invariants for 
the transformation groups in all the four classes.
\end{abstract}

\section{Introduction}\label{rudra:Intro}

\indent
\indent The position of contact
transformations~\cite{rudra:Eisenhart,rudra:Bluman,rudra:Leach,rudra:Anderson}
lies in between the point transformations and the Lie-B\"{a}cklund 
transformations~\cite{rudra:Bluman,rudra:Kyria}. However, in studying the 
dynamics of a system the group of contact transformations has
a very important position. Because of the continuity 
conditions in quantum mechanics of the wavefunction and its 
space derivatives, groups of contact transformations play 
such important roles in the dynamics of physical systems.

Point transformation groups for any set of dif\/ferential
equations involve transformations among the independent 
space-time variables and the dependent variables. The 
generators of the transformation group thus involve only 
these variables. The groups of contact transformations, on the 
other hand, involve these variables as well as the gradients
of the dependent variables. The contact relations connecting 
the gradients with the original variables are thus included 
in the set of dif\/ferential equations. If the generators do not 
involve the gradients in an essential manner, then the contact 
transformation group is {\it not\/} essentially dif\/ferent 
from the point transformation group~\cite{rudra:Bluman}.

In a previous work~\cite{rudra:Rudra} we have studied symmetry
classif\/ication of two-body central potentials for the point 
transformation groups of time-dependent Schr\"{o}dinger 
equation for a two-body system. Here, we have done a similar 
study of the groups of contact transformations of the
same system. What we have found here, and what could not be 
known without this detailed analysis, is that the groups of 
contact transformations for this system are {\it not\/} 
essentially dif\/ferent from the corresponding groups of point 
transformations. Thus we have {\it four\/} classes of
two-body central potentials:
\begin{enumerate}
\item a constant potential with a 31-parameter Lie group,
\item a harmonic oscillator potential with a 20-parameter Lie group,
\item an inverse square {\it potential\/} with a 16-parameter Lie 
group, and  
\item all other potentials with a 14-parameter Lie group. 
\end{enumerate}

Using Lie-Jacobi's method~\cite{rudra:Anderson,rudra:Kyria} we have 
calculated the functional bases of invariants for all these 
cases. Any invariant of these groups can be functionally 
expressed in terms of these base invariants. For the
constant potential case there are 5 base invariants. The 
corresponding number for all the other cases is 4.

\section{Contact transformations and Schr\"{o}dinger equation}
\label{rudra:Contact}

 The method~\cite{rudra:Bluman} for obtaining the group of 
contact transformation is a generalization of the method of 
extended group~\cite{rudra:Eisenhart,rudra:Bluman} for obtaining that 
of point transformations. We give below the essential points 
as it appears in our system of 2 particles of masses $m_1$ and
$m_2$ at positions ${\vec{r}}_1$ and ${\vec{r}}_2$.

We use the relative space coordinate $\vec{r} = {\vec{r}}_1 - 
{\vec{r}}_2$, the centre-of-mass space coordinate $\vec{R} = 
(m_1{\vec{r}}_1 + m_2{\vec{r}}_2)/(m_1 + m_2)$, the reduced 
mass $m = m_1m_2/(m_1 + m_2)$ and the total mass
$M = m_1 + m_2$. It is to be noted that $0 < m/M \leq 1/4$.
The limit $m/M=1/4$ occurs when the two particles have equal
masses. Positronium atom and a homonuclear diatomic molecule are
two important physical systems which have this limiting value of $m/M$.
The other limiting value $m/M=0$ occurs when one of the masses
and hence $M$ is $\infty$. This is same as ignoring the motion
of the centre-of-mass. This does {\it not\/} describe a true two-body 
system and cannot thus throw any light on the classif\/ication
of inter-particle potentials.

In terms of these variables, the 2-particle Schr\"{o}dinger 
equation for the wavefunction ${\Psi}^0$ becomes
\begin{equation}
{\Delta}^0 \equiv i{\Psi}^0_{q^0}+\frac{\hbar}{2M}\sum_{\alpha}
{\Psi}^{c\alpha}_{q^{c\alpha}} + \frac{\hbar}{2m}\sum_{\lambda}
{\Psi}^{r\lambda}_{q^{r\lambda}} - v(r)\Psi^0 = 0,  \label{rudra:eq^1} 
\end{equation}
with the contact conditions
\be
{\Delta}^{c\alpha} \equiv {\Psi}^0_{q^{c\alpha}} - 
{\Psi}^{c\alpha}  =  0 , \label{rudra:eq^2a} 
\ee
\be
{\Delta}^{r\lambda} \equiv {\Psi}^0_{q^{r\lambda}} - 
{\Psi}^{r\lambda}  =  0 . \label{rudra:eq^2b}
\ee
We have written the physicist's version of the equation, keeping
the Planck's constant, the relevant masses and the imaginary
number $\imath$.
We use a compact notation $q^a$, $a = 0$, $c\alpha$, $r\lambda$, where 
$q^0 = t$, $q^{c\alpha} = R_{\alpha}$, $q^{r\lambda} = r_{\lambda}$. 
Here and later  on $\alpha$, $\lambda$ {\it etc.\/} will mean 
the cartesian components, subscripted variables (other than the 
cartesian components) will mean derivatives with respect to those 
subscripts, the letters $c$ and $r$ will mean center-of-mass 
and relative coordinates, and $t$ will mean time.

Equations (\ref{rudra:eq^2a}), (\ref{rudra:eq^2b}) are actually the 
def\/initions of the gradient variables. The inter-particle 
potential $v(r)$ has been taken to be of central nature. 
In terms of the gradient variables the Schr\"{o}dinger
equation is now of the f\/irst order.

The generators of the Lie group transformations in the space 
of $q^a$, ${\Psi}^a$ are of the form
\begin{equation}
X = \sum_{a} \left[{\xi}^a(q,\Psi) \frac{\partial}{\partial q^a} 
+ {\chi}^a(q,\Psi) \frac{\partial}{\partial {\Psi}^a}\right]  ,
\label{rudra:eq^3}
\end{equation} 
with $a = 0$, $c\alpha$, $r\lambda$. The arguments of ${\xi}^a$ and
${\chi}^a$ contain the collection of all $q^a$ and ${\Psi}^a$. In 
the Racah nomenclature the $\xi$s and $\chi$s are called the 
velocity vectors of the generator~$X$. If ${\chi}^0$ explicitly 
contains the gradients ${\Psi}^{c\alpha}$, ${\Psi}^{r\lambda}$, 
then the contact transformation is essentially dif\/ferent from a point 
transformation \cite{rudra:Bluman}. 

As in the case of point transformations, the f\/irst extention of $X$ 
is written as
\begin{equation}
X^{(1)} = X + \sum_{a,b} {\chi}^{a;b} \frac{\partial}{\partial 
{\Psi}^{a}_{q^b}} .  \label{rudra:eq^4} 
\end{equation}
Here,
\begin{equation}
{\chi}^{a;b} = {\chi}^{a}_{q^b} - \sum_{b'} {\Psi}^{a}_{q^{b'}} 
{\xi}^{b'}_{q^b} - \sum_{a',b'} {\Psi}^{a'}_{q^b} 
{\Psi}^{a}_{q^{b'}} {\xi}^{b'}_{{\Psi}^{a'}} + \sum_{a'} 
{\Psi}^{a'}_{q^b} {\chi}^{a}_{{\Psi}^{a'}} .  \label{rudra:eq^6}
\end{equation}

The ef\/fect of $X^{(1)}$ on the dif\/ferent $\Delta$s of 
equations (\ref{rudra:eq^1}), (\ref{rudra:eq^2a}), (\ref{rudra:eq^2b}) are    
\begin{equation}
X^{(1)} {\Delta}^0 \equiv i {\chi}^{0;t} 
 + \frac{\hbar}{2M} \sum_{\alpha} {\chi}^{c\alpha , c\alpha}
 + \frac{\hbar}{2m} \sum_{\lambda} {\chi}^{r\lambda , r\lambda}
 - v(r){\chi}^0 - \frac{v(r)'}{r} \sum_{\lambda} {\xi}^{r\lambda} 
   r_{\lambda} \Psi  =  0  , \label{rudra:eq^5a} 
\end{equation}
\be
X^{(1)} {\Delta}^{c\alpha} \equiv {\chi}^{0;c\alpha} - 
{\chi}^{c\alpha}  =  0 ,
\label{rudra:eq^5b}  
\ee
\be
X^{(1)} {\Delta}^{r\lambda} \equiv {\chi}^{0;r\lambda} - 
{\chi}^{r\lambda}  =  0 .  \label{rudra:eq^5c}
\ee

\section{Def\/ining equations of group generators}\label{rudra:Define}

 The group of contact transformations for the 
equation (\ref{rudra:eq^1}) will be uniquely known when the velocity 
vectors ${\xi}^a$ and ${\chi}^a$ will be obtained. These 
velocity vectors satisfy an overcomplete set of dif\/ferential 
equations known as the def\/ining equations. These def\/ining
equations are obtained by separately equating to zero the 
coef\/f\/icients of the dif\/ferent monomials in ${\Psi}^{a}_{q^b}$ 
appearing in equations (\ref{rudra:eq^5a}), (\ref{rudra:eq^5b}), (\ref{rudra:eq^5c}).

{\advance\topsep-3pt

In our case we get the def\/ining relations:
\be
{\xi}^{0}_{{\Psi}^{c\alpha}} \equiv {\xi}^{0}_{{\Psi}^{r\lambda}}
  =  0, \label{rudra:eq^7a}  
\ee
\be
{\xi}^{0}_{q^{c\alpha}} + {\Psi}^{c\alpha} {\xi}^{0}_{{\Psi}^0}
  \equiv  {\xi}^{0}_{q^{r\lambda}} + {\Psi}^{r\lambda}
 {\xi}^{0}_{{\Psi}^0} =  0,  \label{rudra:eq^7b}  
\ee
\be
{\xi}^{c\alpha}_{{\Psi}^{r\lambda}} \equiv 
{\xi}^{r\lambda}_{{\Psi}^{c\alpha}}  =  0,   \label{rudra:eq^7c} 
\ee
\be
\frac{1}{M} {\delta}_{\alpha\beta} {\xi}^{r\lambda}_{{\Psi}^{r\mu}}
+ \frac{1}{m} {\delta}_{\lambda\mu} {\xi}^{c\alpha}_{{\Psi}^{c\beta}}
       =  0,  \label{rudra:eq^7d}  
\ee
\be
{\delta}_{\nu\lambda} {\xi}^{r\sigma}_{{\Psi}^{r\mu}} +
 {\delta}_{\sigma\mu} {\xi}^{r\nu}_{{\Psi}^{r\lambda}} 
      =  0,                \label{rudra:eq^7e}   
\ee
\be
{\delta}_{\gamma\alpha} {\xi}^{c\sigma}_{{\Psi}^{c\beta}} +
 {\delta}_{\sigma\beta} {\xi}^{c\gamma}_{{\Psi}^{c\alpha}}
   =  0,                \label{rudra:eq^7f}  
\ee
\be
{\chi}^{0}_{{\Psi}^{c\alpha}} - 
 \sum_{\beta} {\Psi}^{c\beta} {\xi}^{c\beta}_{{\Psi}^{c\alpha}} - 
 \sum_{\mu} {\Psi}^{r\mu} {\xi}^{r\mu}_{{\Psi}^{c\alpha}}
       =  0,      \label{rudra:eq^7g}    
\ee
\be
{\chi}^{0}_{{\Psi}^{r\lambda}} -
 \sum_{\beta} {\Psi}^{c\beta} {\xi}^{c\beta}_{{\Psi}^{r\lambda}} - 
 \sum_{\mu} {\Psi}^{r\mu} {\xi}^{r\mu}_{{\Psi}^{r\lambda}}
    =  0,      \label{rudra:eq^7h}  
\ee
\begin{equation} 
\ba{l}
\ds  \left[{\chi}^{0}_{q^{c\alpha}} - 
 \sum_{\beta} {\Psi}^{c\beta} {\xi}^{c\beta}_{q^{c\alpha}} -
 \sum_{\mu} {\Psi}^{r\mu} {\xi}^{r\mu}_{q^{c\alpha}}\right] 
\vspace{3mm}\\
\ds \qquad + 
{\Psi}^{c\alpha} \left[{\chi}^{0}_{{\Psi}^0}  - 
 \sum_{\beta} {\Psi}^{c\beta} {\xi}^{c\beta}_{{\Psi}^0} -
 \sum_{\mu} {\Psi}^{r\mu} {\xi}^{r\mu}_{{\Psi}^0}\right] - 
 {\chi}^{c\alpha}  =  0,  
\ea \label{rudra:eq^7i}   
\end{equation}
\begin{equation}  
\ba{l}
\ds   \left[{\chi}^{0}_{q^{r\lambda}} - 
 \sum_{\beta} {\Psi}^{c\beta} {\xi}^{c\beta}_{q^{r\lambda}} - 
 \sum_{\mu} {\Psi}^{r\mu} {\xi}^{r\mu}_{q^{r\lambda}}\right] 
\vspace{3mm}\\
\ds \qquad +
{\Psi}^{r\lambda} \left[{\chi}^{0}_{{\Psi}^0} - 
 \sum_{\beta} {\Psi}^{c\beta} {\xi}^{c\beta}_{{\Psi}^0} -
 \sum_{\mu} {\Psi}^{r\mu} {\xi}^{r\mu}_{{\Psi}^0}\right] - 
{\chi}^{r\lambda}  =  0,   
\ea \label{rudra:eq^7j} 
\end{equation}  
\begin{equation}
{\delta}_{\alpha\gamma}\left[{\chi}^{0}_{{\Psi}^0} - 
\sum_{\beta} {\Psi}^{c\beta} {\xi}^{c\beta}_{{\Psi}^0} - 
\sum_{\mu} {\Psi}^{r\mu} {\xi}^{r\mu}_{{\Psi}^0} - 
{\xi}^{0}_{q^0}\right] + {\xi}^{c\gamma}_{q^{c\alpha}} -  
{\chi}^{c\gamma}_{{\Psi}^{c\alpha}} + 
{\Psi}^{c\alpha} {\xi}^{c\gamma}_{{\Psi}^0}   =  0, \label{rudra:eq^7k}  
\end{equation}              
\begin{equation}
{\delta}_{\lambda\nu}\left[{\chi}^{0}_{{\Psi}^0} - 
\sum_{\beta} {\Psi}^{c\beta} {\xi}^{c\beta}_{{\Psi}^0} - 
\sum_{\mu} {\Psi}^{r\mu} {\xi}^{r\mu}_{{\Psi}^0} - 
{\xi}^{0}_{q^0}\right] + {\xi}^{r\nu}_{q^{r\lambda}} - 
{\chi}^{r\nu}_{{\Psi}^{r\lambda}} + 
{\Psi}^{r\lambda} {\xi}^{r\nu}_{{\Psi}^0}   =  0, \label{rudra:eq^7l}  
\end{equation}
\be
\frac{\hbar}{2M} {\chi}^{c\alpha}_{{\Psi}^{r\lambda}} -  
\frac{\hbar}{2m} \left[{\xi}^{c\alpha}_{q^{r\lambda}} + 
{\Psi}^{r\lambda} {\xi}^{c\alpha}_{{\Psi}^0}\right]  
 =  0, \label{rudra:eq^7m}   
\ee
\be
\frac{\hbar}{2m} {\chi}^{r\lambda}_{{\Psi}^{c\alpha}} - 
\frac{\hbar}{2M} \left[{\xi}^{r\lambda}_{q^{c\alpha}} +  
{\Psi}^{c\alpha} {\xi}^{r\lambda}_{{\Psi}^0}\right]  
 =  0, \label{rudra:eq^7n} 
\ee
\be
\ba{l}
\ds i\left[{\chi}^{0}_{q^0} - 
\sum_{\beta} {\Psi}^{c\beta} {\xi}^{c\beta}_{q^0} -  
\sum_{\mu} {\Psi}^{r\mu} {\xi}^{r\mu}_{q^0}\right] + 
v(r){\Psi}^0\left[{\chi}^{0}_{{\Psi}^0} - 
\sum_{\beta} {\Psi}^{c\beta} {\xi}^{c\beta}_{{\Psi}^0} -  
\sum_{\mu} {\Psi}^{r\mu} {\xi}^{r\mu}_{{\Psi}^0}\right] 
\vspace{3mm}  \\    
\ds \qquad +\frac{\hbar}{2M} \sum_{\alpha}
\left[{\chi}^{c\alpha}_{q^{c\alpha}} + 
{\Psi}^{c\alpha} {\chi}^{c\alpha}_{{\Psi}^0}\right]  +  
\frac{\hbar}{2m}\sum_{\lambda} \left[{\chi}^{r\lambda}_{q^{r\lambda}} 
+ {\Psi}^{r\lambda} {\chi}^{r\lambda}_{{\Psi}^0}\right] 
 \vspace{3mm} \\   
\ds \qquad -v(r)\left[{\chi}^0 + {\Psi}^0 {\xi}^{0}_{q^0} +  
\frac{v(r)'}{rv(r)} {\Psi}^0 \sum_{\lambda} r_{\lambda} 
{\xi}^{r\lambda}\right]  =  0.  
\ea     \label{rudra:eq^7o} 
\ee
We have obtained the general solution of these def\/ining equations
and have indicated the method in the Appendix 1. The velocity
vectors are expressed in terms of the auxilliary functions
\be
\ba{l}
\ds F^0(t,\vec{R},\vec{r})  =  A_0(t) -
\frac{iM}{\hbar} \sum_{\alpha} f^{c\alpha}_0(t)' R_\alpha - 
\frac{im}{\hbar} \sum_{\lambda} f^{r\lambda}_0(t)' r_\lambda  
\vspace{3mm}  \\
\ds \qquad + \frac{iM}{4\hbar} b(t)'' \sum_{\alpha} R_{\alpha}^2 + 
\frac{im}{4\hbar} b(t)'' \sum_{\lambda} r_{\lambda}^2,
\ea \label{rudra:eq^8a}    
\ee
\be
\ds F^{c\alpha}(t,\vec{R},\vec{r})  =  f_{0}^{c\alpha}(t) -
\frac{1}{2} b(t)' R_{\alpha} + 
\sum_{\beta\gamma} e_{\alpha\beta\gamma}f_{1}^{c\gamma}R_{\beta}+ 
m \sum_{\lambda} f_{0}^{c\alpha;r\lambda} r_{\lambda},
\label{rudra:eq^8b}    
\ee
\be
F^{r\lambda}(t,\vec{R},\vec{r})  =  f_{0}^{r\lambda}(t) - 
\frac{1}{2} b(t)' r_{\lambda} + 
\sum_{\mu\nu} e_{\lambda\mu\nu} f_{1}^{r\nu} r_{\mu} - 
M \sum_{\alpha} f_{0}^{c\alpha;r\nu} R_{\alpha},
\label{rudra:eq^8c}
\ee
and are of the form
\be
{\xi}^{0}(t)   =  b(t),  
\ee
\be
{\xi}^{c\alpha}(t,\vec{R},\vec{r})   =  -F^{c\alpha}(t.\vec{R},
\vec{r}), \label{rudra:eq^9a}  
\ee
\be
{\xi}^{r\lambda}(t,\vec{R},\vec{r})   =  -F^{r\lambda}(t,\vec{R},
\vec{r}), \label{rudra:eq^9b}  
\ee
\be
{\chi}^{0}(t,\vec{R},\vec{r},{\Psi}^0)   =  f^{0}(t,\vec{R},
\vec{r})+  {\Psi}^0 F^{0}(t,\vec{R},\vec{r}),  \label{rudra:eq^9c}  
\ee
\be
\ba{l}
\ds {\chi}^{c\alpha}(t,\vec{R},\vec{r},{\Psi}^a)   = 
\frac{{\partial}{\chi}^{0}(t,\vec{R},\vec{r},{\Psi}^0)}{{\partial}
R_{\alpha}}-  M \sum_{\lambda} {\Psi}^{r\lambda} 
f_{0}^{c\alpha;r\lambda} \vspace{3mm}\\
\ds \qquad - \sum_{\beta} {\Psi}^{c\beta} \left[ \frac{1}{2} 
{\delta}_{\alpha\beta} b(t)' - {\delta}_{\alpha\beta} 
F^{0}(t,\vec{R},\vec{r}) +  \sum_{\gamma} e_{\alpha\beta\gamma} 
f_{1}^{c\gamma} \right] ,   
\ea \label{rudra:eq^9d}  
\ee
\be
\ba{l}
\ds {\chi}^{r\lambda}(t,\vec{R},\vec{r},{\Psi}^a)   =    
\frac{{\partial}{\chi}^{0}(t,\vec{R},\vec{r},{\Psi}^0)}{{\partial}
r_{\lambda}} +m\sum_{\alpha} {\Psi}^{c\alpha}
f_{0}^{c\alpha;r\lambda} 
\vspace{3mm}\\
\ds \qquad -  \sum_{\mu} {\Psi}^{r\mu} \left[ \frac{1}{2} 
{\delta}_{\lambda\mu} b(t)' - {\delta}_{\lambda\mu} 
F^{0}(t,\vec{R},\vec{r})+\sum_{\nu} e_{\lambda\mu\nu}f_{1}^{r\nu} 
\right],   
\ea \label{rudra:eq^9e}
\ee
where $F^{0}(t,\vec{R},\vec{r})$ and $f^{0}(t,\vec{R},\vec{r})$ 
satisfy 
\be
\ba{l}
\ds i \frac{{\partial}F^{0}(t,\vec{R},\vec{r})}{{\partial} t} +  
\frac{\hbar}{2M} \sum_{\alpha} \frac{{\partial}^2 F^{0}(t,\vec{R},
\vec{r})}{({\partial}R_\alpha)^2}+\frac{\hbar}{2m} \sum_{\lambda} 
\frac{{\partial}^2F^{0}(t,\vec{R},\vec{r})}{({\partial}r_\lambda)^2}
  \vspace{3mm}  \\  
\ds \qquad - v(r)b(t)' + \frac{v(r)'}{r} \sum_{\lambda}  r_\lambda   
F^{r\lambda}(t,\vec{R},\vec{r}) =  0,
\ea     \label{rudra:eq^10a} 
\ee
\be
\ba{l}
\ds i \frac{{\partial}f^{0}(t,\vec{R},\vec{r})}{{\partial}t} +  
\frac{\hbar}{2M}  \sum_{\alpha} \frac{{\partial}^2 
f^{0}(t,\vec{R},\vec{r})}{({\partial}R_\alpha)^2} + 
\frac{\hbar}{2m} \sum_{\lambda} \frac{{\partial}^2 
f^{0}(t,\vec{R},\vec{r})}{({\partial}r_\lambda)^2}   
\vspace{3mm}  \\  
\ds \qquad - v(r)f^{0}(t,\vec{R},\vec{r}) =  0. 
\ea   \label{rudra:eq^10b}
\ee

} %%% end {\advance\topsep-3pt}

{\small

\begin{center}
\begin{tabular}{|c|l|} \hline
& \\[-3.5mm]
$\ba{c} \mbox{Name}/\\
\mbox{Symbol of the generator} \ea $ & Form of the generator \\
& \\[-3.5mm]
 \hline
& \\[-3.5mm]
$\begin{array}{c} \mbox{\sl Scaling}:\\ X_{S} \end{array}$ 
&  $\ds {\Psi}^0 
\frac{\partial}{{\partial}{{\Psi}^0}} + \sum_{\alpha} {\Psi}^{c\alpha} 
\frac{\partial}{{\partial} {{\Psi}^{c\alpha}}} + \sum_{\lambda} 
{\Psi}^{r\lambda} \frac{\partial}{{\partial}{{\Psi}^{r\lambda}}}$ \\
& \\[-3.5mm]
\hline
& \\[-3.5mm]
$\begin{array}{c} \mbox{\sl time~translation}:\\X^t \end{array}$ & 
$\ds i\frac{\partial}{{\partial}t}$ \\
& \\[-3.5mm]
\hline
& \\[-3.5mm]
$\begin{array}{c} \mbox{\sl centre~of~mass~coordinate}\\
\mbox{\sl space~~translations}:\\ 
X_{T}^{c\alpha}\end{array}$ & $\ds -i \frac{\partial}{{\partial}
{R_{\alpha}}}$ \\ 
& \\[-3.5mm]
\hline
& \\[-3.5mm]
$\begin{array}{c} \mbox{\sl relative~coordinate}\\ 
\mbox{\sl space~translations}:\\
X_{T}^{r\lambda}\end{array}$ & 
$\ds  -i \frac{\partial}{{\partial} {r_{\lambda}}}$ \\ 
& \\[-3.5mm]
\hline
& \\[-3.5mm]
$\begin{array}{c} \mbox{\sl centre of mass coordinate}\\ 
\mbox{\sl Galilean transformations}:\\ 
X_{G}^{c\alpha}\end{array}$ & $\ds tX_{T}^{c\alpha} + 
 \frac{M}{\hbar} \left[R_{\alpha}X_S+{\Psi}^0\frac{\partial}{{\partial}
{\Psi}^{c\alpha}}\right]$ \\ 
& \\[-3.5mm]
\hline
& \\[-3.5mm]
$\begin{array}{c} \mbox{\sl relative~coordinate}\\
\mbox{\sl Galilean transformations}:
\\X_{G}^{r\lambda}\end{array}$ & $\ds tX_{T}^{r\lambda} + 
 \frac{m}{\hbar} \left[r_{\lambda} X_{S} + {\Psi}^0 \frac{\partial}{{\partial}
 {\Psi}^{r\lambda}} \right]$ \\  
& \\[-3.5mm]
\hline  
& \\[-3.5mm]
$\begin{array}{c} \mbox{\sl centre of mass coordinate}\\
\mbox{\sl space rotations}:\\
X_{R}^{c\alpha}\end{array}$ & $\ds -i\sum_{\beta\gamma}e_{\alpha
\beta\gamma}\left[R_{\beta}\frac{\partial}{{\partial}R_{\gamma}}+{\Psi}^{c
 \beta}\frac{\partial}{{\partial}{\Psi}^{c\gamma}} \right]$ \\ 
& \\[-3.5mm]
\hline
& \\[-3.5mm]
$\begin{array}{c} \mbox{\sl relative coordinate}\\
\mbox{\sl space rotations}:\\
X_{R}^{r\lambda}\end{array}$ & $\ds -i\sum_{\mu\nu}e_{\lambda\mu\nu}
\left[r_{\mu}\frac{\partial}{{\partial}r_{\nu}}+{\Psi}^{r\mu}\frac{\partial}{
 {\partial}{\Psi}^{r\nu}} \right]$ \\ 
& \\[-3.5mm]
\hline  
& \\[-3.5mm]
$\begin{array}{c} \mbox{\sl cross-rotations}:\\
X^{c\alpha;r\lambda}\end{array}$
 & $\ds  -i\sqrt{\frac{m}{M}} \left[r_{\lambda}\frac{\partial}{{\partial}
 R_{\alpha}} - {\Psi}^{c\alpha}\frac{\partial}{{\partial}{\Psi}^{r\lambda}} 
 \right]+i\sqrt{\frac{M}{m}}\left[R_{\alpha}\frac{\partial}{{\partial}
 r_{\lambda}}-{\Psi}^{r\lambda}\frac{\partial}{{\partial}{\Psi}^{c\alpha}} 
 \right]$ \\  
& \\[-3.5mm]
\hline
& \\[-3.5mm]
$X_1$ & $\ds 2tX^t - \sum_{\alpha} R_{\alpha}X_{T}^{c\alpha}- \sum_{\lambda} 
 r_{\lambda}X_{T}^{r\lambda}+2v_0 tX_S + i{\Psi}^0 \frac{\partial}{
 {\partial}{\Psi}^0}$ \\  
& \\[-3.5mm]
\hline
& \\[-3.5mm]
$X_2$ &  $\ds t^2X^t - t \sum_{\alpha} R_{\alpha} X_{T}^{c\alpha} -
 t \sum_{\lambda} r_{\lambda} X_{T}^{r\lambda}$\\
& $\qquad \ds - \left[\frac{M}{2\hbar}\sum_{\alpha}
R_{\alpha}^2+\frac{m}{2\hbar} 
 \sum_{\lambda} r_{\lambda}^2 +4it -v_0t^2\right]X_S$ \\    
 & $\ds \qquad -{\Psi}^0\left[-it\frac{\partial}{{\partial}{\Psi}^0}+
\frac{M}{\hbar}\sum_{\alpha} R_{\alpha}\frac{\partial}{{\partial}
{\Psi}^{c\alpha}} + \frac{m}{\hbar} \sum_{\lambda} r_{\lambda}
\frac{\partial}{{\partial}{\Psi}^{r\lambda}}\right]$ \\
& \\[-3.5mm]
 \hline
& \\[-3.5mm]
$\begin{array}{c} \mbox{\sl relative vibrations}:\\ 
X^{r\lambda}_{V,(\pm)}
\end{array}$ & $\ds e^{{\pm i\omega t}}\left[-i\frac{\partial}{{\partial}
r_{\lambda}} \pm \frac{im\omega}{\hbar}r_{\lambda}X_S \pm \frac{im
\omega}{\hbar}{\Psi}^0 \frac{\partial}{{\partial}{\Psi}^{r\lambda}}\right]$ 
\\[3.5mm]
 \hline
\end{tabular}
\end{center}

\noindent
Table 1. Generators that describe the dif\/ferent Lie Algebras of the
 four classes of inter-particle potential. 

}
\bigskip

\noindent
Here, a prime on the function of a single variable denotes
derivative with respect to that variable, $e_{\alpha\beta\gamma}$
and $e_{\lambda\mu\nu}$ are the permutation symbols and
$f_{1}^{c\alpha}$, $f_{1}^{r\lambda}$, $f_{0}^{c\alpha;r\lambda}$ are
constants. Since ${\chi}^{0}(t,\vec{R},\vec{r},{\Psi}^0)$ does
not contain ${\Psi}^{c\alpha}$ and ${\Psi}^{r\lambda}$, the
group of contact transformations for this system is not
essentially dif\/ferent from that of point transformations.
However, the generators now contain derivatives with respect
to the gradient variables, because the group of contact
transformations is the f\/irst extension of the group of point
transformations.

Equation (\ref{rudra:eq^10b}) is nothing but the original
Schr\"{o}dinger equation and $f^{0}(t,\vec{R},\vec{r})$ is a
solution of equation (\ref{rudra:eq^1}). This is the symmetry
corresponding to linear superposition principle of the
Schr\"{o}dinger equation and forms an inf\/inite dimensional
invariant subgroup of the total group. The factor group
modulo this subgroup is the physical group of interest and
will be referred to as the group of contact transformations.

Substituting $F^{0}(t,\vec{R},\vec{r})$ from 
equation (\ref{rudra:eq^8a}) in equation (\ref{rudra:eq^10a}) and equating
coef\/f\/icients of dif\/ferent monomials in the space coordinates
to zero, we get
\be
b(t)'''   =   0,    \label{rudra:eq^11a}  
\ee
\be
f_{0}^{c\alpha}(t)''   =  0,  \label{rudra:eq^11b}  
\ee
\be
\frac{m}{\hbar} f_{0}^{r\lambda}(t)'' +  
\frac{v(r)'}{r} f_{0}^{r\lambda}(t)   =   0,   \label{rudra:eq^11c} 
\ee
\be
iA_{0}(t)' + \frac{3i}{2} b(t)'' - 
\left[ v(r)+ \frac{1}{2} rv(r)' \right] b(t)'   =   0, 
\label{rudra:eq^11d}
\ee
and  either
\begin{equation}
  v(r)' = 0   \label{rudra:eq^11e}
\end{equation}
or  
\begin{equation}
 f_{0}^{c\alpha;r\lambda} = 0.  \label{rudra:eq^11f}
\end{equation}

\section{Symmetry classif\/ication of interparticle potentials}\label{rudra:Class}

>From the solution of equations (\ref{rudra:eq^11a})--(\ref{rudra:eq^11f}) we get
four dif\/ferent classes of interparticle potentials. 
This complete symmetry analysis of the 2-particle Schr\"{o}dinger
equation as far as the dynamics of the system is concerned 
shows that contact transformations do not enforce any more 
restriction than that already required on the basis of point 
transformation symmetry of the system. Their group algebras are the
f\/irst extensions of the algebras of point transformation symmetries
of the corresponding potentials~\cite{rudra:Rudra}.

These group algebras are described in terms of the generators given
in Table~1.

The letters $T$, $G$, $R$ and $V$ in the symbols of the generators
denote  space translational, galilean, space rotational and vibrational modes
described by these generators. We have kept the imaginary number 
$\imath$ in the forms of the generators so that these can be identif\/ied 
with the usual quantum mechanical operators for energy, linear and 
angular momenta.

In actual calculations with the group algebras, the structure constants
(commutation relations in physicists' parlance) are of the greatest
help. They are given in Appendix~2.

Any Lie Algebra $L$ is a semi-direct product of its Radical $R$ and
a semisimple part $L/R$. The semisimple part $L/R$ is again a direct
sum of ideals which, as subalgebras, are simple
\cite{rudra:Naimark,rudra:Humphreys}. These characteristics of the Lie Algebras
for the dif\/ferent classes of inter-particle potentials are given in
Table~2.

It is to be noted that the generators for the space translational and
Galilean as well as vibrational transformations always belong to the 
Radical of the Lie Algebra.

{
\small 

\begin{center}
\begin{tabular}{|c|c|c|c|c|} \hline 
& & & & \\[-3.5mm]
\mbox{\sl Inter-particle} & \mbox{\sl Constant} & \mbox{\sl Harmonic}
& \mbox{\sl Inverse square} 
 & \mbox{\sl Arbitrary} \\
\mbox{\sl potential}: &   & \mbox{\sl oscillator}  &   &    \\
$v(r)$ & $=v_0$ & $\ds =v_0+\frac{m{\omega}^2r^2}{2\hbar}$ & 
 $\ds =v_0-\frac{v_1}{r^2}$ & $\neq v_0$, \\
 &   &   &   & $\ds v_0+\frac{m{\omega}^2r^2}{2\hbar}$,\\[3.5mm]
 &  &  &  &  $\ds v_0-\frac{v_1}{r^2}$ \\ 
& & & & \\[-3.5mm]\hline
& & & & \\[-3.5mm]
Generators of & \{$X_S$,$X^t$,$X^{c\alpha}_{G,T.R}$, & \{$X_S$,$X^t$,
 $X^{c\alpha}_{G,T,R}$, & \{$X_S$,$X^t$,$X^{c\alpha}_{G,T,R}$, & \{$X_S$,
 $X^t$, \\
 Lie Algebra, & $X^{r\lambda}_{G,T,R}$,$X^{c\alpha ;r\lambda}$,
 & $X^{r\lambda}_R$,$X^{r\lambda}_{V,(\pm )}$\} & $X^{r\lambda}_R$,
 $X_1$,$X_2$\} & $X^{c\alpha}_{G,T,R}$,$X^{r\lambda}_R$\} \\ 
$L$ & $X_1$,$X_2$\} &  &  &    \\ 
& & & & \\[-3.5mm]
\hline 
& & & & \\[-3.5mm]
Dimension & 31 & 20 & 16 & 14 \\ 
& & & & \\[-3.5mm]
\hline 
& & & & \\[-3.5mm]
Solvability, & None & None & None & None \\
Nilpotency, &   &   &   &   \\
Simplicity, &   &   &   &   \\
Semisimplicity &   &   &   &   \\  
& & & & \\[-3.5mm]
\hline
& & & & \\[-3.5mm]
Centre, $Z(L)$ & \{$X_S$\} & \{$X_S$\} & \{$X_S$\} & \{$X_S$\} \\ 
& & & & \\[-3.5mm]
\hline
& & & & \\[-3.5mm]
Radical, $R$ & $\{X_S,X^{c\alpha}_{G,T},X^{r\lambda}_{G,T}\}$
 & $\{X_S,X^{c\alpha}_{G,T},X^{r\lambda}_{V,(\pm)}\}$
 & $\{X_S,X^{c\alpha}_{G,T}\}$  & 
$\{X_S,X^{c\alpha}_{G,T}\}$  \\
& & & & \\[-3.5mm]
\hline 
& & & & \\[-3.5mm]
Semisimple  & $\!I_1=\{X^t,X_1,X_2\},\!$ & 
$I_1=\{X^t\}$, & $\!I_1=\{X^t,X_1,X_2\}$,
 & $I_1=\{X^t\}$, \\
part as  & $I_2=\{X^{c\alpha}_R,X^{r\lambda}_R$, 
 & $I_2=\{X^{c\alpha}_R\}$, 
& $I_2=\{X^{c\alpha}_R\}$,
 & $I_2=\{X^{c\alpha}_R\}$,  \\
direct sum &  $X^{c\alpha;r\lambda}\}$ & $I_3=\{X^{r\lambda}_R \}$ & 
 $I_3=\{X^{r\lambda}_R\}$ & $I_3=\{X^{r\lambda}_R\}$ \\ 
$L/R={\sum_i}^{\oplus} I_i$ &  &   &  &  \\
of simple  &   &   &   &    \\ 
ideals &   &   &   &   \\  
& & & & \\[-3.5mm]
\hline
& & & & \\[-3.5mm]
Cartan  & \{$X_S$,$X^{c3}_R$,$X^{r3}_R$, & \{$X_S$,$X^t$,$X^{c3}_{G,T,R}$,
 & \{$X_S$,$X^{c3}_R$,$X^{r3}$, & \{$X_S$,$X^t$, \\
subalgebra & $X^{c3;r3}$,$X_1$\} & $X^{r3}_R$\} & $X_1$\},
 & $X^{c3}_{G,T,R}$,$X^{r3}_R$\} \\  
$H$ &   &   &   &  \\[1mm]    \hline
\end{tabular} 
\end{center}

\noindent
Table 2. Characteristics of the Lie Algebras of the Symmetry groups
of the dif\/ferent classes of inter-particle potential.

}

\bigskip

We note that the Radical, being solvable, has only 1-dimensional
irreducible representations (irreps)~\cite{rudra:Naimark}. Thus the
irreps of $L$ are obtained if the irreps of $L/R$ are known. To this
end we give in Table~3 the algebraic characteristics of the
simple subalgebras appearing in the direct sum of $L/R$ for the
dif\/ferent classes of inter-particle potential.

\section{Functional bases of invariants}  \label{rudra:Base}

In order to investigate integrability of a
dynamical system we require the functional bases of invariants
in terms of which all invariants of the dynamical system
can be expressed functionally. These base invariants of a Lie algebra
{\sl L} generate the {\sl Centre} of the Universal Enveloping algebra
of {\sl L}. Since they commute with all the generators of {\sl L},
they appear as conserved quantities of the system. Lie's
method~\cite{rudra:Lie,rudra:Goursat,rudra:Racah,rudra:Abellanas} has been utilized to obtain
the base invariants for the four symmetry groups of Section~\ref{rudra:Class}. 
If the functional base has $s$ invariants $I_1,I_2,\ldots,I_s$, each 
a function of the $r$ generators $X_a$ of the symmetry group, then
\[
\left[X_a,I_b\right] = 0,  \qquad a=1,\ldots,r, \quad b=1,\ldots,s.
\]

{
\small

\begin{center}
\begin{tabular}{|c|c|c|c|c|}  \hline
&&&& \\[-3.5mm]
Algebra: $L$ & $L_1$ & $L_2$ & $L_3$ & $L_4$ \\  
&&&& \\[-3.5mm]
\hline 
&&&& \\[-3.5mm]
Generators: & \{$X^t$,$X_1$,$X_2$\} & \{$X^{c\alpha}_R$, $X^{r\lambda}_R$,
$X^{c\alpha ;r\lambda}$\} & \{$X^{\alpha}_R$\} & \{$X^t$\} \\
\{$X_i$\} &  &   &   &    \\ 
&&&& \\[-3.5mm]
\hline
&&&& \\[-3.5mm]
Dimension & 3 & 15 & 3 & 1 \\ 
&&&& \\[-3.5mm]
\hline
&&&& \\[-3.5mm]
Cartan &   &   &   &   \\
subalgebra: & $\ds\left\{ -\frac{i}{2}X_1\right\}$ & \{$X^{c3}_R$,$X^{r3}_R$,
 $X^{c3;r3}$\} & \{$X^3_R$\} & \{$X^t$\} \\
$H$ &   &   &   &  \\ 
&&&& \\[-3.5mm]
\hline
&&&& \\[-3.5mm]
Rank & 1 & 3 & 1 & 1 \\
&&&& \\[-3.5mm]
 \hline
&&&& \\[-3.5mm]
${\begin{array}{c} \mbox{Base of}\\{\rm simple}\\{\rm roots:}\\ \Delta
 \end{array}}$ & $\alpha = 1$ & ${\alpha}_1=\left({\begin{array}{r} 
 -1\\ 0\\ -1 \end{array}}\right)$, ${\alpha}_2=\left({\begin{array}{r}
  -1\\ 0\\ 1 \end{array}}\right)$, & $\alpha$ = 1 & void \\
 &   & ${\alpha}_3=\left({\begin{array}{r} 1\\ 1\\ 0 \end{array}}\right)$ 
 &   &  \\ 
&&&& \\[-3.5mm]
\hline 
&&&& \\[-3.5mm]
${\begin{array}{c} {\rm Root}\\{\rm system}\end{array}}$ & $\pm\alpha$ &
${\begin{array}{c} \pm{\alpha}_1,\pm{\alpha}_2,\pm{\alpha}_3, \\
 \pm\left({\alpha}_1+{\alpha}_3\right),\pm\left({\alpha}_2+{\alpha}_3
 \right), \\ \pm\left({\alpha}_1+{\alpha}_2+{\alpha}_3\right) \end{array}}$ 
 & $\pm\alpha$ & void \\  
&&&& \\[-3.5mm]
\hline
&&&& \\[-3.5mm]
${\begin{array}{c} {\rm Isomorphy}\\{\rm to} \end{array}}$ & $O(3)$ &
 $O(6)$ & $O(3)$ & $U(1)$ \\  
&&&& \\[-3.5mm]
\hline
&&&& \\[-3.5mm]
${\begin{array}{c} {\rm Standard}\\{\rm set of}\\{\rm generators}
 \end{array}}$ & 
${\begin{array}{c} E_{\alpha}=X_2 \\E_{-\alpha}=X^t
 \end{array}}$ & ${\begin{array}{c} E_{\pm{\alpha}_1}=\left(X^{c1}_R \mp
 iX^{c2}_R\right)\\[1mm] +\left(X^{c1;r3}\mp iX^{c2;r3}\right),\\[1mm] 
 E_{\pm{\alpha}_2}=\left(X^{c1}_R\mp iX^{c2}_R\right) \\[1mm] -\left(
 X^{c1;r3}\mp iX^{c2;r3}\right),\\[1mm] 
E_{\pm{\alpha}_3}=\left(X^{c1;r1}
 -X^{c2;r2}\right)\\[1mm] \pm i\left(X^{c1;r2}+X^{c2;r1}\right),\\[1mm]
 E_{\pm\left({\alpha}_1+{\alpha}_3\right)}=\left(X^{r1}_R\pm iX^{r2}_R
 \right)\\[1mm] -\left(X^{c3;r1}\pm iX^{c3;r2}\right), \\[1mm]
 E_{\pm\left(
 {\alpha}_2+{\alpha}_3\right)}=\left(X^{r1}_R\pm iX^{r2}_R\right) \\[1mm]
 +\left(X^{c3;r1}\pm iX^{c3;r2}\right), \\[1mm]
E_{\pm\left({\alpha}_1
 +{\alpha}_2+{\alpha}_3\right)}=\left(X^{c1;r1}+X^{c2;r2}\right) \\[1mm]
 \pm i\left(X^{c1;r2}-X^{c2;r1}\right)\end{array}}$ 
 & ${\begin{array}{c} E_{\pm\alpha}= \\[1mm] X^1_R\pm iX^2_R \end{array}}$
 & void \\[3.5mm]  \hline
\end{tabular}
\end{center}

\centerline{Table 3. Characteristics of the simple subalgebras forming the
semisimple parts in Table~2.}
}

\bigskip

\noindent
If $I$ is any other invariant so that
\[
\left[X_a,I\right] =0, \qquad a=1,\ldots,r,
\]
then $I$ can be functionally expressed as
\[
 I \equiv I(I_1,\ldots,I_s).
\]
If the base has  the constant as its only member, then
the system is completely chaotic. If on the other hand
$s=r$ and all the generators give mutually commuting
invariants, then the system is fully integrable. Actual
dynamical systems are almost always in between these two
extreme cases. In the four symmetry classes obtained in
Section~\ref{rudra:Class}, the same thing happens.

In Table~4 we def\/ine auxilliary operators in terms 
of which the functional bases of invariants for the four 
classes of distinct inter-particle potentials are given in
Table~5.

{
\small

\begin{center}
\begin{tabular}{|c|l|} \hline
& \\[-3.5mm]
$\begin{array}{c} {\rm Inter-particle}\\{\rm potential}\end{array}$ &
  Auxilliary operators  \\
& \\[-3.5mm]   \hline
& \\[-3.5mm]
$\begin{array}{c} \mbox{\sl Constant}\\\mbox{\sl potential}\end{array}$ &
$\begin{array}{l} 
\ds Y^{c\alpha}_R=X_SX^{c\alpha}_R+\frac{\hbar}{M}
 \sum_{\beta\gamma} e_{\alpha\beta\gamma}X^{c\beta}_TX^{c\gamma}_G, \\
\ds  Y^{r\lambda}_R=X_SX^{r\lambda}_R+\frac{\hbar}{m}\sum_{\mu\nu}
 e_{\lambda\mu\nu}X^{r\mu}_TX^{r\nu}_G, \\
\ds  Y^{c\alpha ;r\lambda}=X_SX^{c\alpha ;r\lambda}+\frac{\hbar}{\sqrt{mM}}
\left[X^{c\alpha}_GX^{r\lambda}_T-X^{c\alpha}_T X^{r\lambda}_G
\right], \vspace{1mm}\\ 
 \ds Y^t=X_SX^t+v_0\left(X_S\right)^2+\frac{\hbar}{2M}\sum_{\alpha} 
 \left(X^{c\alpha}_T\right)^2+\frac{\hbar}{2m}\sum_{\lambda} \left(
 X^{r\lambda}_T\right)^2,\\
\ds Y_1=X_SX_1 -4i\left(X_S\right)^2
 +\frac{\hbar}{M}\sum_{\alpha} X^{c\alpha}_TX^{c\alpha}_G
 +\frac{\hbar}{m}\sum_{\lambda} X^{r\lambda}_TX^{r\lambda}_G,\\
 \ds Y_2=X_SX_2+\frac{\hbar}{2M}\sum_{\alpha} \left(X^{c\alpha}_G
 \right)^2+\frac{\hbar}{2m}\sum_{\lambda} \left(X^{r\lambda}_G
 \right)^2 \end{array}$ \\ 
& \\[-3.5mm]
\hline
& \\[-3.5mm]
$\begin{array}{c} \mbox{\sl Harmonic}\\ 
\mbox{\sl oscillator}\\ \mbox{\sl potential}
 \end{array}$ & $\begin{array}{l} 
\ds Y^{c\alpha}_R=X_SX^{c\alpha}_R
 +\frac{\hbar}{M}\sum_{\beta\gamma} e_{\alpha\beta\gamma}X^{c\beta}_T
 X^{c\gamma}_G,\\
\ds Y^{r\lambda}_R=X_SX^{r\lambda}_R+
 \frac{i\hbar}{2m\omega}\sum_{\mu\nu} e_{\lambda\mu\nu}
 X_{V,(+)}^{r\mu}X_{V,(-)}^{r\nu},\\
\ds Y^t=X_SX^t+\frac{\hbar}{2M}
 \sum_{\alpha} \left(X_{T}^{c\alpha}\right)^2+\frac{\hbar}{2m}
 \sum_{\lambda} X_{V,(+)}^{r\lambda}X_{V,(-)}^{r\lambda}\end{array}$  
 \\  
& \\[-3.5mm]
\hline  
& \\[-3.5mm]
$\begin{array}{c} \mbox{\sl Inverse}\\
\mbox{\sl square}\\
\mbox{\sl potential}\end{array}$  &
$\begin{array}{l} 
\ds Y_{R}^{c\alpha}=X_SX_{R}^{c\alpha}+\frac{\hbar}{M}
\sum_{\beta\gamma} e_{\alpha\beta\gamma}X_{T}^{c\beta}
X_{G}^{c\gamma},\\ 
\ds Y^t=X_SX^t+v_0\left(X_S\right)^2+\frac{\hbar}{2M}\sum_{\alpha} \left(
X_{T}^{c\alpha}\right)^2,\\
\ds Y_1=X_SX_1-4i\left(X_S\right)^2
 +\frac{\hbar}{M}\sum_{\alpha} X_{T}^{c\alpha}X_{G}^{c\alpha},\\ 
\ds Y_2=X_SX_2+ \frac{\hbar}{2M} \sum_{\alpha} \left(X_{G}^{c\alpha}
\right)^2 \end{array}$   \\ 
& \\[-3.5mm]
\hline
& \\[-3.5mm]
 $\begin{array}{c} \mbox{\sl Arbitrary}\\
\mbox{\sl  potential}\end{array}$ &
$\begin{array}{l} 
\ds Y_{R}^{c\alpha}=X_SX_{R}^{c\alpha}+\frac{\hbar}{M} 
\sum_{\beta\gamma} e_{\alpha\beta\gamma}X_{T}^{c\beta}X_{G}^{c\gamma}, 
 \\ 
\ds Y^t=X_SX^t+\frac{\hbar}{2M}\sum_{\alpha} \left(X_{T}^{c\alpha}
\right)^2 \end{array}$ \\[4.5mm]  \hline
\end{tabular}
\end{center}

\centerline{Table 4. Auxilliary operators for the four distinct classes of
 inter-particle potentials.}
}

\bigskip

  In all these cases $I_S$ is the scaling operator and 
$I^t$ is essentially the energy operator, $I_{R}^c$ and $I_{R}^r$ are 
the c.m.coordinate and the relative coordinate angular momentum           
operators. Except for the case of constant potential, these are
the only invariants. The only potential that has more invariants
is the constant potential and the two extra invariants $I^{c;r}_3$
and $I^{c;r}_4$ of degrees 3 and 4 in the generators make this
apparently simple system actually more complicated yet more
systematic.

{\small

\begin{center}
\begin{tabular}{|c|l|} \hline
& \\[-3.5mm]
$\begin{array}{c} {\rm Inter-particle}\\{\rm potential}\end{array}$ &
 Base Invariants \\ 
& \\[-3.5mm]
\hline
& \\[-3.5mm]
$\begin{array}{c} \mbox{\sl Constant}\\
\mbox{\sl potential}\end{array}$ &
  $\begin{array}{l} I_S=X_S,\\ 
\ds I^t=Y_2Y^t-
 \frac{1}{4}\left(Y_1\right)^2,\vspace{1mm}\\
\ds I^0_R=\sum_{\alpha} \left(Y^{c\alpha}_R
 \right)^2+\sum_{\lambda} \left(Y^{r\lambda}_R\right)^2
 +\sum_{\alpha\lambda} \left(Y^{c\alpha ;r\lambda}\right)^2, \vspace{1mm}\\
\ds  I^{c;r}_3 = \sum_{\alpha\lambda} Y^{c\alpha}_RY^{c\alpha ;r\lambda}
 Y^{r\lambda}_R  \vspace{1mm}\\
\ds \qquad          -\frac{1}{6}\sum_{\alpha\beta\gamma}
 \sum_{\lambda\mu\nu} e_{\alpha\beta\gamma}e_{\lambda\mu\nu}
 Y^{c\alpha ;r\lambda}Y^{c\beta ;r\mu}Y^{c\gamma ;r\nu} , \vspace{1mm}\\
\ds I^{c;r}_4 = 
 \left[\sum_{\alpha} \left(Y^{c\alpha}_R\right)^2\right]\left[\sum_{\lambda}
 \left(Y^{r\lambda}_R\right)^2\right]\\         
\ds \qquad +\sum_{\alpha} \left[
 \sum_{\lambda} Y^{c\alpha ;r\lambda}Y^{r\lambda}_R\right]^2+\sum_{\lambda} 
 \left[\sum_{\alpha} Y^{c\alpha}_RY^{c\alpha ;r\lambda}\right]^2\\
    \ds    \qquad    +\frac{1}{4}\sum_{\alpha\lambda} \left[\sum_{\beta\gamma}
 \sum_{\mu\nu} e_{\alpha\beta\gamma}e_{\lambda\mu\nu}Y^{c\beta ;r\mu}
 Y^{c\gamma ;r\nu} \right]^2\\
\ds          \qquad -\sum_{\alpha\lambda} Y^{c\alpha}_R
 \left[\sum_{\beta\gamma} \sum_{\mu\nu} e_{\alpha\beta\gamma}e_{\lambda\mu\nu}
Y^{c\beta ;r\mu}Y^{c\gamma ;r\nu}\right] Y^{r\lambda}_R
\end{array}$\\ 
& \\[-3.5mm]
\hline
& \\[-3.5mm]
$\begin{array}{c} \mbox{\sl Harmonic}\\
\mbox{\sl oscillator}\\
\mbox{\sl potential}
 \end{array}$ & $\begin{array}{l} I_S = X_S,\\I^t = Y^t,\vspace{1mm}\\
\ds  I_{R}^c = \sum_{\alpha} \left(Y_{R}^{c\alpha}\right)^2,
\vspace{1mm}\\
\ds  I_{R}^r = \sum_{\lambda} \left(Y_{R}^{r\lambda}\right)^2 \end{array}$ 
 \\ 
& \\[-3.5mm]
 \hline  
& \\[-3.5mm]
$\begin{array}{c} \mbox{\sl Inverse}\\
\mbox{\sl square}\\
\mbox{\sl potential}\end{array}$  &
 $\begin{array}{l} I_S = X_S,\\
\ds I^t = Y_2Y^t-\frac{1}{4}\left(Y_1\right)^2, \vspace{1mm}\\
 \ds I_{R}^c = \sum_{\alpha} \left(Y_{R}^{c\alpha}\right)^2,\vspace{1mm}\\
 \ds I_{R}^r = \sum_{\lambda} \left(X_{R}^{r\lambda}\right)^2 \end{array}$  
 \\ 
& \\[-3.5mm]
\hline
& \\[-3.5mm]
$\begin{array}{c} \mbox{\sl Arbitrary}\\
\mbox{\sl  potential}\end{array}$ &
 $\begin{array}{l} I_S = X_S,\\I^t = Y^t,\\
\ds I_{R}^c = \sum_{\alpha} 
 \left(Y_{R}^{c\alpha}\right)^2, \vspace{1mm}\\
\ds I_{R}^r = \sum_{\lambda} 
 \left(X_{R}^{r\lambda}\right)^2\end{array}$ \\[3.5mm]   \hline
\end{tabular}
\end{center}

\centerline{Table 5. Functional base of invariants for the dif\/ferent classes
of inter-particle potentials.} 
}

\bigskip

It is to be noted that in all the expressions in this section
involving products of operators symmetrized forms have to be
taken.

\medskip

{\bf Acknowledgement:} I thank the referee for suggesting a number of
improvements to the 
manuscript.

\section*{Appendix 1} 

 In this Appendix we indicate how the def\/ining 
equations (\ref{rudra:eq^7a})--(\ref{rudra:eq^7o}) are solved to obtain the
group generators.

Equations (\ref{rudra:eq^7a}), (\ref{rudra:eq^7b}) give ${\xi}^0 \equiv 
{\xi}^{0}(t)$. From equations (\ref{rudra:eq^7c})--(\ref{rudra:eq^7f}) it 
follows that
\begin{equation}
{\xi}^{c\alpha}\equiv {\xi}^{c\alpha}(t,\vec{R},\vec{r},{\Psi}^0),
\qquad {\xi}^{r\lambda}\equiv {\xi}^{r\lambda}(t,\vec{R},\vec{r},
{\Psi}^0).       \label{rudra:A3}
\end{equation}
Equations (\ref{rudra:eq^7g}), (\ref{rudra:eq^7h}) thus give
\begin{equation}
{\chi}^0 \equiv {\chi}^{0}(t,\vec{R},\vec{r},{\Psi}^0). \label{rudra:A4}
\end{equation}
Using the auxilliary function
\begin{equation}
F(q^a,{\Psi}^a) = {\chi}^0-\sum_{\alpha} {\Psi}^{c\alpha}
{\xi}^{c\alpha}-\sum_{\lambda} {\Psi}^{r\lambda}
{\xi}^{r\lambda},           \label{rudra:A5}
\end{equation}
equations (\ref{rudra:eq^7i}) and (\ref{rudra:eq^7j}) now become
\be
{\xi}^{c\alpha}  =  -\frac{{\partial}F}{{\partial}{\Psi}^{c
\alpha}}, \label{rudra:A6a} 
\ee
\be
{\xi}^{r\lambda}  =  -\frac{{\partial}F}{{\partial}{\Psi}^{r
\lambda}}, \label{rudra:A6b} 
\ee
\be
{\chi}^{c\alpha}  =  \frac{{\partial}F}{{\partial}q^{c\alpha}} + 
{\Psi}^{c\alpha}\frac{{\partial}F}{{\partial}{\Psi}^0},  
\label{rudra:A6c} 
\ee
\be
{\chi}^{r\lambda}  =  \frac{{\partial}F}{{\partial}q^{r\lambda}} + 
{\Psi}^{r\lambda}\frac{{\partial}F}{{\partial}{\Psi}^0}, 
\label{rudra:A6d} 
\ee
\be
{\chi}^0  =  F-\sum_{\alpha} {\Psi}^{c\alpha}\frac{{\partial}F}{
{\partial}{\Psi}^{c\alpha}}-\sum_{\lambda} {\Psi}^{r\lambda}
\frac{{\partial}F}{{\partial}{\Psi}^{r\lambda}}.  \label{rudra:A6e}
\ee
Here $F$ is linear in ${\Psi}^{c\alpha}$ and ${\Psi}^{r\lambda}$. 
Because of this linearity, equations (\ref{rudra:eq^7i})--(\ref{rudra:eq^7l}) 
give
\begin{equation}
{\xi}^{c\alpha} \equiv {\xi}^{c\alpha}(t,\vec{R},\vec{r}),
\qquad 
{\xi}^{r\lambda} \equiv {\xi}^{r\lambda}(t,\vec{R},\vec{r}), 
\label{rudra:A7}
\end{equation}
and equations (\ref{rudra:A5}), (\ref{rudra:A6c}), (\ref{rudra:A6d}) and (\ref{rudra:A7}) 
imply linearity of ${\chi}^{c\alpha}$ and ${\chi}^{r\lambda}$ with 
respect to ${\Psi}^{c\alpha}$ and ${\Psi}^{r\lambda}$.
Equations (\ref{rudra:eq^7k}), (\ref{rudra:eq^7l}) together with equations
(\ref{rudra:A5})--(\ref{rudra:A6d}) now give
\be
\ba{l}
{\xi}^{c\alpha}_{q^{c\beta}}+{\xi}^{c\beta}_{q^{c\alpha}}  =  
{\delta}_{\alpha\beta}{\xi}_{q^0}^0,  \vspace{2mm}\\
{\xi}^{r\lambda}_{q^{r\mu}}+{\xi}^{r\mu}_{q^{r\lambda}}  = 
{\delta}_{\lambda\mu}{\xi}_{q^0}^0,  
\ea\label{rudra:A8}
\ee
and equations (\ref{rudra:eq^7m}), (\ref{rudra:eq^7n}) become
\begin{equation}
m{\xi}^{r\lambda}_{q^{c\alpha}}+M{\xi}^{c\alpha}_{q^{r\lambda}}  
= 0.                      \label{rudra:A9}
\end{equation}

From equation (\ref{rudra:eq^7o}) equating separately to zero terms
quadratic in, linear in and independent of ${\Psi}^{c\alpha}$
and ${\Psi}^{r\lambda}$, we get
\be
{\chi}_{{\Psi}^0,{\Psi}^0}^0  =  0,  \label{rudra:A10a} 
\ee
\be
i{\xi}^{c\alpha}_{q^0}  =  -\frac{\hbar}{2M} \sum_{\beta} 
{\xi}^{c\alpha}_{q^{c\beta},q^{c\beta}} -\frac{\hbar}{2m}
\sum_{\mu} {\xi}^{c\alpha}_{q^{r\mu},q^{r\mu}}+\frac{\hbar}{M}
{\chi}_{{\Psi}^0,q^{c\alpha}}^0,  \label{rudra:A10b} 
\ee
\be
i{\xi}^{r\lambda}_{q^0}  =  -\frac{\hbar}{2M} \sum_{\beta} 
{\xi}^{r\lambda}_{q^{c\beta},q^{c\beta}} -\frac{\hbar}{2m}
\sum_{\mu} {\xi}^{r\lambda}_{q^{r\mu},q^{r\mu}}+\frac{\hbar}{m}
{\chi}_{{\Psi}^0,q^{c\lambda}}^0,  \label{rudra:A10c} 
\ee
\be
\ba{l}
\ds i{\chi}_{q^0}^0  =  -\frac{\hbar}{2M}\sum_{\alpha} 
{\chi}_{q^{c\alpha},q^{c\alpha}}^0 - \frac{\hbar}{2m}\sum_{\lambda} 
{\chi}_{q^{r\lambda},q^{r\lambda}}^0+v(r){\chi}^0 
\vspace{3mm}\\
\ds \qquad + v(r){\Psi}^0\left[{\xi}_{q^0}^0-{\chi}_{{\Psi}^0}^0 +
\frac{v(r)'}{rv(r)}\sum_{\lambda} r_{\lambda}{\xi}^{r\lambda}
\right].   
\ea  \label{rudra:A10d}
\ee
These will result in
\be
\ba{l}
\ds {\xi}^0  =  b(t), \qquad {\xi}^{c\alpha} = -F^{c\alpha}(t,\vec{R},
\vec{r}), \qquad {\xi}^{r\lambda} = -F^{r\lambda}(t,\vec{R},\vec{r}),
\vspace{3mm} \\
\ds {\chi}^0  =  f^0(t,\vec{R},\vec{r})+{\Psi}^0F^0(t,\vec{R},
\vec{r}),  \vspace{3mm} \\
\ds {\chi}^{c\alpha}  =  {\chi}_{R_\alpha}^0+\sum_{\lambda} 
{\Psi}^{r\lambda}F_{R_\alpha}^{r\lambda}+ 
\sum_{\beta} {\Psi}^{c\beta}\left[F^0{\delta}_{\alpha\beta}
+F^{c\beta}_{R_\alpha}\right],  \vspace{3mm} \\
\ds {\chi}^{r\lambda}  =  {\chi}_{r\lambda}^0+\sum_{\alpha} 
{\Psi}^{c\alpha}F^{c\alpha}_{r_\lambda} +
 \sum_{\mu} {\Psi}^{r\mu}\left[F^0{\delta}_{\lambda\mu}
+ F^{r\mu}_{r_\lambda}\right],  
\ea \label{rudra:A11}
\ee
with
\be
F^{c\alpha}_{R_\beta}+F^{c\beta}_{R_\alpha}  =  -b(t)'
{\delta}_{\alpha\beta},   \label{rudra:A12.1}  
\ee
\be
F^{r\lambda}_{r_\mu}+F^{r\mu}_{r_\lambda}  =  -b(t)'
{\delta}_{\lambda\mu}, \label{rudra:A12.2} 
\ee
\be
MF^{c\alpha}_{r_\lambda}+mF^{r\lambda}_{R_\alpha}  =  0, 
\label{rudra:A12.3}
\ee
and
\be
iF^{c\alpha}_t  =  -\frac{\hbar}{M}F_{R_\alpha}^0 -  
  \frac{\hbar}{2M}\sum_{\beta}F^{c\alpha}_{R_\beta,R_\beta} - 
\frac{\hbar}{2m}\sum_{\mu} F^{c\alpha}_{r_\mu,r_\mu}, 
\label{rudra:A13.1} 
\ee
\be
iF^{r\lambda}_t  =  -\frac{\hbar}{m}F_{r_\lambda}^0- 
  \frac{\hbar}{2M}\sum_{\beta} F^{r\lambda}_{R_\beta,R_\beta} -
\frac{\hbar}{2m}\sum_{\mu} F^{r\lambda}_{r_\mu,r_\mu}, 
\label{rudra:A13.2} 
\ee
\be
iF^0_t  =  v(r)b(t)'-\frac{v(r)'}{r}\sum_{\lambda} 
r_{\lambda}F^{r\lambda}- 
   \frac{\hbar}{2M}\sum_{\alpha} F_{R_\alpha,R_\alpha}^0 -
\frac{\hbar}{2m}\sum_{\lambda} F_{r_\lambda,r_\lambda}^0, 
\label{rudra:A14.1} 
\ee
\be
if^0_t  =  -\frac{\hbar}{2M}\sum_{\alpha} f_{R_\alpha,R_\alpha}^0-
\frac{\hbar}{2m}\sum_{\lambda} f_{r_\lambda,r_\lambda}^0+v(r)f^0.      
\label{rudra:A14.2}
\ee
Equations (\ref{rudra:A14.1}) and (\ref{rudra:A14.2}) are obtained from the terms
that are linear in and independent of ${\Psi}^0$ in equation
(\ref{rudra:A10d}). Equation (\ref{rudra:A12.1}) gives us the result that 
$F^{c\alpha}$ is linear in $\vec{R}$; and equation (\ref{rudra:A12.2}) 
shows that $F^{r\lambda}$ is linear in $\vec{r}$. Equation 
(\ref{rudra:A12.3}) will now give $F^0$, $F^{c\alpha}$ and $F^{r\lambda}$
as in equations (\ref{rudra:eq^8a})--(\ref{rudra:eq^8c}) and ${\chi}^{c\alpha}$,
${\chi}^{r\lambda}$ as in equations (\ref{rudra:eq^9d}), (\ref{rudra:eq^9e}), the
dif\/ference is that $f_1^{c\gamma}$, $f_1^{r\nu}$ and $f_0^{c\alpha;
r\lambda}$ are yet functions of time $t$. Equations (\ref{rudra:A13.1}),
(\ref{rudra:A13.2}) now ensure that these are indeed independent of $t$.

From equation (\ref{rudra:A14.1}) we also obtain 
equations (\ref{rudra:eq^11a})--(\ref{rudra:eq^11f}).

\section*{Appendix 2} 

In this Appendix we give the structure constants of the generators
given in Table~1.
\[
\left[X^t,X_{G}^{c\alpha}\right]=iX_{T}^{c\alpha}, \qquad  \left[
X^t,X_{G}^{r\lambda}\right] = iX_{T}^{r\lambda},
\qquad 
\left[X^t,X_1\right] = 2i\left(X^t+v_0X_S\right),
\]
\[
 \left[
X^t,X_2\right] = iX_1+4X_S, 
\qquad \left[X_{T}^{c\alpha},X_{G}^{c\beta}\right]=-\frac{iM}{\hbar}
\delta_{\alpha\beta}X_S,
\]
\[
\left[X_{T}^{c\alpha},X_{R}^{c\beta}
\right]=i\sum_{\gamma} e_{\alpha\beta\gamma}X_{T}^{c\gamma}, 
\qquad \left[X_{T}^{c\alpha},X_1\right] =iX_{T}^{c\alpha}, \qquad \left[
X_{T}^{c\alpha},X_2\right]=iX_{G}^{c\alpha}, 
\]
\[
\left[X_{T}^{c\alpha},X^{c\beta;r\lambda}\right] =i\sqrt{\frac{M}{m}}
\delta_{\alpha\beta}X_{T}^{r\lambda}, 
\qquad
\left[X_{G}^{c\alpha},X_{R}^{c\beta}\right] =i\sum_{\gamma}  
e_{\alpha\beta\gamma}X_{G}^{c\gamma}, 
\]
\[
 \left[X_{G}^{c\alpha},X_1
\right]=-iX_{G}^{c\alpha}, 
\qquad 
\left[X_{G}^{c\alpha},X^{c\beta;r\lambda}\right]=i\sqrt{\frac{M}{m}}
\delta_{\alpha\beta}X_{G}^{r\lambda}, 
\]
\[
\left[X_{R}^{c\alpha},X_{R}^{c\beta}\right]=i\sum_{\gamma}  
e_{\alpha\beta\gamma}X_{R}^{c\gamma}, \qquad \left[X_{R}^{c\alpha},
X^{c\beta;r\lambda}\right]=i\sum_{\gamma} e_{\alpha\beta\gamma}
X^{c\gamma;r\lambda},  
\]
\[
\left[X_{T}^{r\lambda},X_{G}^{r\mu}\right]=-\frac{im}{\hbar}
\delta_{\lambda\mu}X_S, \qquad \left[X_{T}^{r\lambda},X_{R}^{r\mu}
\right]=i\sum_{\nu} e_{\lambda\mu\nu}X_{T}^{r\nu}, 
\]
\[
\left[X_{T}^{r\lambda},X_1\right]=iX_{T}^{r\lambda}, \qquad
\left[X_{T}^{r\lambda},X_2\right]=iX_{G}^{r\lambda}, 
\qquad 
\left[X_{T}^{r\lambda},X^{c\alpha;r\mu}\right]=-i\sqrt{\frac{m}{M}}
\delta_{\lambda\mu}X_{T}^{c\alpha}, 
\]
\[
\left[X_{G}^{r\lambda},X_{R}^{r\mu}\right]=i\sum_{\nu} 
e_{\lambda\mu\nu}X_{G}^{r\nu}, \qquad \left[X_{G}^{r\lambda},X_1
\right]=-iX_{G}^{r\lambda}, 
\]
\[
\left[X_{G}^{r\lambda},X^{c\alpha;r\mu}\right]=-\sqrt{\frac{m}{M}} 
\delta_{\lambda\mu}X_{G}^{c\alpha}, 
\]
\[
\left[X_{R}^{r\lambda},X_{R}^{r\mu}\right]=i\sum_{\nu} 
e_{\lambda\mu\nu}X_{R}^{r\nu}, \qquad  \left[X_{R}^{r\lambda},
X^{c\alpha;r\mu}\right]=i\sum_{\nu} e_{\lambda\mu\nu} 
X^{c\alpha;r\nu},  
\]
\[
\left[X^{c\alpha;r\lambda},X^{c\beta;r\mu}\right]  =  i
\left(\delta_{\alpha\beta}\sum_{\nu} e_{\lambda\mu\nu}X_{R}^{r\nu} 
 +\delta_{\lambda\mu}\sum_{\gamma} e_{\alpha\beta\gamma}
X_{R}^{c\gamma}\right),  
\]
\[
\left[X_1,X_2\right]  =  2iX_2,
\qquad
\left[X^t,X^{r\lambda}_{V,(\pm)}\right]={\mp}{\omega}X^{r\lambda}_{
V,(\pm)}, 
\]
\[
  \left[X_{R}^{r\lambda},X^{r\mu}_{V,(\pm)}\right]=i
\sum_{\nu} e_{\lambda\mu\nu}X^{r\nu}_{V,(\pm)}, 
\qquad
\left[X^{r\lambda}_{V,(+)},X^{r\mu}_{V,(-)}\right]=-\frac{2m
\omega}{\hbar}\delta_{\lambda\mu}X_S.   \label{rudra:eq^15}
\]

\label{rudra-lp}

\end{document}